\documentclass[prl,twocolumn, showpacs,amssymb,superscriptaddress]{revtex4}

\usepackage{graphicx, color}

\newcommand{\ket}[1]{\ensuremath{|#1\rangle}}
\newcommand{\bra}[1]{\ensuremath{\langle #1 |}}
\newcommand{\proj}[1]{\ket{#1}\!\bra{#1}}

\begin{document}
\title{Experimental approximation of the Jones polynomial with one quantum bit}

\author{G. Passante}
\affiliation{Institute for Quantum Computing and Dept. of Physics, University of Waterloo, Waterloo, ON, N2L 3G1, Canada.}

\author{O. Moussa}
\affiliation{Institute for Quantum Computing and Dept. of Physics, University of Waterloo, Waterloo, ON, N2L 3G1, Canada.}

\author{C.A. Ryan }
\affiliation{Institute for Quantum Computing and Dept. of Physics, University of Waterloo, Waterloo, ON, N2L 3G1, Canada.}

\author{R. Laflamme}
\affiliation{Institute for Quantum Computing and Dept. of Physics, University of Waterloo, Waterloo, ON, N2L 3G1, Canada.}
\affiliation{Perimeter Institute for Theoretical Physics, Waterloo, ON, N2J 2W9, Canada}

\date{\today}

\begin{abstract}
We present experimental results approximating the Jones polynomial using 4 qubits in a liquid state nuclear magnetic resonance quantum information processor. This is the first experimental implementation of a complete problem for the DQC1 model of quantum computation, which uses a single qubit accompanied by a register of completely random states.  The Jones polynomial is a knot invariant that is important not only to knot theory, but also to statistical mechanics and quantum field theory.  The implemented algorithm is a modification of the algorithm developed by Shor and Jordan suitable for implementation in NMR.  These experimental results show that for the restricted case of knots whose braid representations have four strands and exactly three crossings, identifying distinct knots is possible $91\%$ of the time.  These results demonstrate the high level of experimental control currently available in liquid state NMR.
\end{abstract}


\maketitle

Quantum information processors have the potential to solve some problems exponentially faster than current classical methods \cite{Nielsen2000Quantum-Computa}.  While much effort has been concentrated on the most conventional circuit model of computation which involves preparation of pure fiducial quantum states, other models of computation, where only one pure quantum bit is required, still offer efficient solutions to classically intractable problems.  Deterministic quantum computation with one quantum bit (DQC1) is such a model \cite{Knill1998Power-of-One-Bi}. It extracts the power of one bit of quantum information alongside a register of many qubits in a completely random state.  Study of DQC1 was originally motivated by liquid state nuclear magnetic resonance (NMR), which is a high temperature ensemble model of quantum computation.
Although this model of computation is weaker than conventional models with many pure qubits, it has been shown to have several important applications where classical methods are inefficient: simulating quantum systems \cite{Knill1998Power-of-One-Bi}, estimating the average fidelity decay under quantum maps \cite{Poulin2004Exponential-Spe}, and quadratically signed weight enumerators \cite{Knill2001Quantum-computi}.  Additionally, the approximation of the Jones polynomial at the fifth root of unity has recently been shown to completely encapsulate the power of DQC1 \cite{Shor2007Estimating-Jone}.  DQC1 algorithms have been experimentally implemented in optics \cite{Lanyon2008Experimental-qu} and liquid and solid state NMR \cite{Ryan2005,Marx2009NMR-Quantum-Cal,Moussa2009} -- none of which has been shown to be DQC1-complete.  In~\cite{Marx2009NMR-Quantum-Cal}, the authors implement a DQC1 algorithm on two qubits to evaluate the Jones polynomial at various points for specific knots.  This letter describes the implementation of an instance of a DQC1-complete algorithm~\cite{Shor2007Estimating-Jone}, which scales for any size knot.

Unlike its name suggests, DQC1 does not require a completely pure qubit to provide an advantage over known classical methods, but rather a small fraction of a pure qubit.  
This pseudo-pure state is almost completely mixed with a small bias towards the ground state, and is used as the control qubit in the DQC1 algorithm.  A unitary is performed on the qubits in the completely mixed state and is controlled by the pseudo-pure qubit.   
\begin{figure}[h]
\includegraphics[scale=0.31]{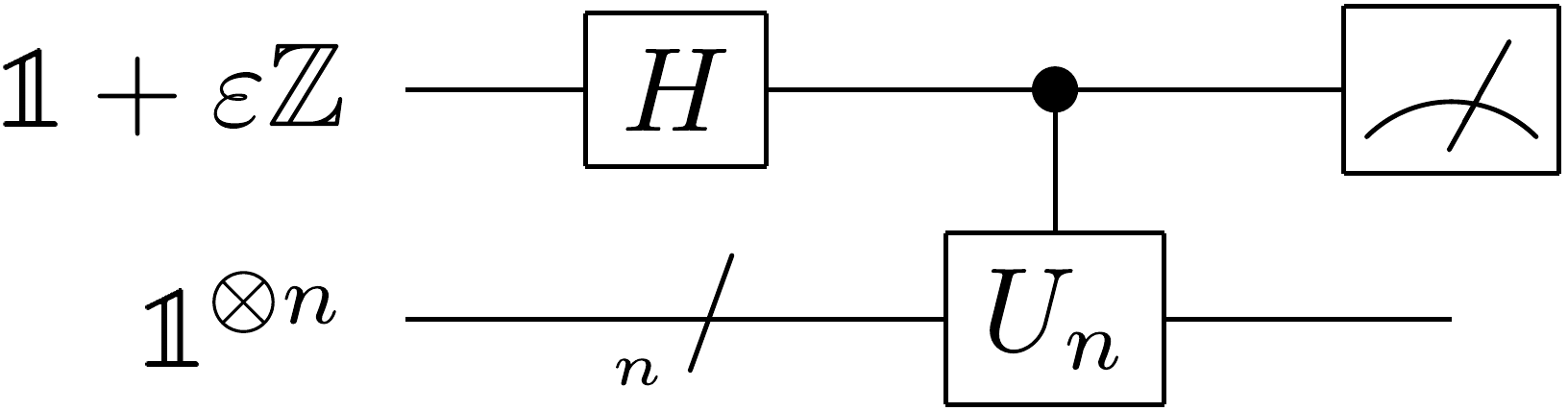}
\caption{The DQC1 circuit where the pure qubit has a bias of $\varepsilon$ towards the ground state.  Measurements of $\langle \sigma_x \rangle$ and $\langle \sigma_y \rangle$ will yield the real and imaginary parts of $\varepsilon\text{Tr}(U_n)/2^n$, respectively.}
\label{dqc1}
\end{figure}
Measurements of $\langle \sigma_x \rangle$ and $\langle \sigma_y \rangle$ yield the real and imaginary parts of the trace of the unitary, normalized by the amount of polarization on the pure qubit.

Applications for the Jones polynomial are extensive in physics; for example, the fields of statistical mechanics, quantum field theory, and quantum gravity would benefit from an efficient method for approximating this polynomial~\cite{Kauffman1991Knots-and-Physi}.  Knot invariants help to solve a fundamental problem in knot theory: determining if two knots, defined as the embedding of the circle in $\mathbb{R}^3$, are topologically different, up to ambient isotopy.  Two knots can only be confirmed identical if one can be maneuvered into the other by a sequence of Reidemeister moves, which keep the topological properties of knots intact.  This process is very tedious as often the sequence of Reidemeister moves require an increase in the number of crossings in the knot.  Even the simplest such problem of identifying the unknot, a circle with no crossings, has been shown to be contained in the complexity class NP~\cite{Hass1999The-computation}.  Knot invariants, such as the Jones polynomial, have the same value for different representations of the same knot.  In other words, if a knot invariant evaluates to different values for two knots, they are guaranteed to be distinct.  This makes them a welcome alternative to sequences of Reidemeister moves.  Unfortunately, exact evaluation of the Jones polynomial at all but a few points is \#P--hard~\cite{Jaeger1990On-the-computat}.  Several efforts for finding quantum algorithms for the Jones polynomial have been attempted and approximations at several special points have been shown to be BQP--complete~\cite{Freedman2002Simulation-of-T,Freedman2002A-Modular-Funct}.  Largely building on this work it was then shown that approximations of the Jones polynomial for trace or plat closures at principal roots of unity can be computed on a quantum computer in polynomial time~\cite{Aharonov2006A-Polynomial-Qu}.  Later it was shown that for the plat closure the problem is BQP-complete~\cite{Aharonov2006The-BQP-hardnes,Wocjan2006The-Jones-Polyn}.  The algorithm developed by Shor and Jordan shows that approximating the Jones polynomial at the fifth root of unity for any knot is a complete problem for DQC1.  

For the purposes of this algorithm, knots are described in the discrete language of braid groups.  Every knot can be written as a braid, which is a series of strands crossings over and under each other with loose ends at both the top and bottom.  Braids can then be converted into a knot by the trace closure which connects the top and bottom ends of the braid in sequential order.  The braid group for $m$ strands $B_m$ is generated by $s_1 \ldots s_{m-1}$ that denote elementary crossings where $s_i$ indicates the $i^{th}$ strand crossing over the $(i+1)^{th}$ strand and $s_i^{-1}$ indicates strand $(i+1)$ crossing over strand $i$.  These elementary crossings satisfy the relations: $s_is_j = s_js_i$ for $|j-i|>1$ and $s_{i+1}s_is_{i+1} = s_is_{i+1}s_i$.  

The implemented algorithm utilizes the Fibonacci representation of the braid group $B_n$, which is described in the context of the Temperley-Lieb recoupling theory~\cite{Kauffman2007q-Deformed-Spin}.  In this theory there are two particles $p$ and $*$, which exhibit the following properties:  $p$ interacts with another $p$ to create a $p$ or a $*$ particle, $*$ interacts with a $p$ to always create a $p$ particle, and two $*$'s never interact.  Strings of these particles create a basis in a complex vector space.  More details of this representation can be found in~\cite{Kauffman2007q-Deformed-Spin}, but for our purposes it suffices to state that for a braid with $m$ strands the basis vectors contain $m+1$ elements of $p$'s and $*$'s with the restriction that no two $*$ particles be beside one another.  These basis vectors are then transformed into the computational basis and unitary matrices $\sigma_i$, which represent each elementary crossing in the braid group, are constructed.  For the particular form of these unitaries, please refer to~\cite{Shor2007Estimating-Jone}.

\begin{figure}[htb]
\includegraphics[scale=0.3]{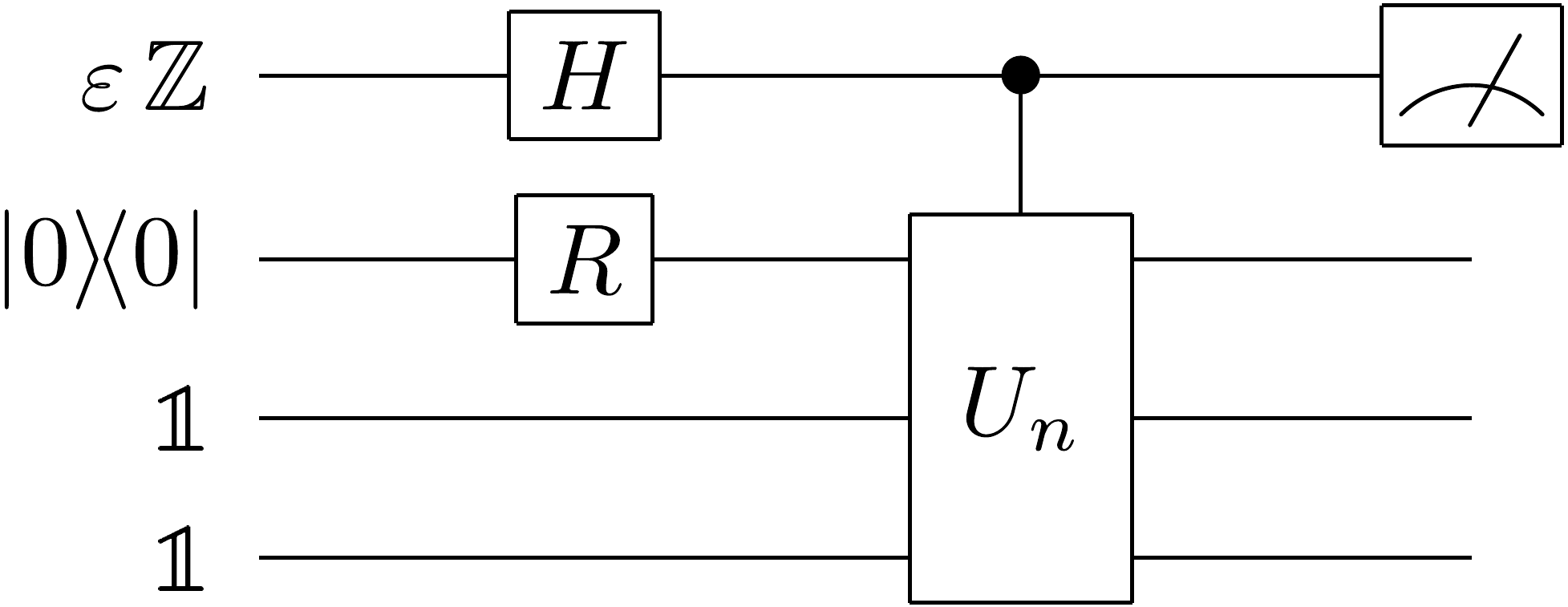}
\caption{Circuit diagram for the approximation of the Jones polynomial for the knots whose braid representations consist of four strands.  The initial state given is the traceless deviation matrix.  The single qubit gates are the Hadamard and the rotation for implementing the weights of the trace.  The measurements performed on the top qubit are expectation values of the Pauli--x and y operators. }
\label{Jones_circuit}
\end{figure}

The algorithm developed by Shor and Jordan approximates the Jones polynomial at the single point $t = e^{2i\pi/5}$ by finding the weighted trace of a unitary that describes the braid representation of the knot.  The algorithm is modified for this implementation and the varied portions are described below.  
The primary difference is in the encoding of the basis states.
The Fibonacci basis vectors consist of four distinct subspaces, only two of which are relevant for the algorithm: the $f_m$ vectors of the form $*\ldots p$ and $f_{m-1}$ of the form $*\ldots *$,  where $f_n= [1,\; 1,\; 2,\; 3,\; \ldots ]$ is the fibonacci sequence.  These are the only two subspaces that are encoded in this implementation.  The Zeckendorf representation, $z^\prime = 2^{n-1}s_1 + \sum_{i=2}^{m-1}s_{i+s_1}f_i$ converts the Fibonacci basis vectors into integers that are then converted to a non-saturated computational basis.  The second notable difference is the method used to calculate the weighted trace, defined as
\begin{eqnarray*}
\text{WTr} \;= &1&\!\!\times(\text{trace of subspace }*\ldots *) + \\
                                 &\phi&\!\!\times(\text{trace of subspace }*\ldots p),
\end{eqnarray*}
where $\phi = (1+\sqrt{5})/2$ is the golden ratio.  Implementing these weights for our encoding is achieved by purifying the second qubit, then applying a rotation taking $\ket{0}$ to $(\sqrt{\phi}\ket{0} + \ket{1})/\sqrt{1+\phi}$, which ensures that each subspace receives the desired weight.  The computational model now contains two initialized qubits, however this modification does not change the computational power as DQC($k$) is known to have the same computational power as DQC1 for $k$ that can grow logarithmically with the total number of qubits~\cite{Shor2007Estimating-Jone}.  The extra basis states are accounted for in the final calculation of the Jones polynomial.  
The circuit for our evaluation of the Jones polynomial for braids with four strands can be seen in figure \ref{Jones_circuit}.
It is worthwhile to note that the off-diagonal elements in the rotated pure qubit do not contribute to the algorithm as the unitary matrices $U_n$ are always block diagonal, thereby eliminating the off-diagonal elements in the calculation of the trace.  The state of the top qubit at the completion of the algorithm is
\[
\rho = \frac{1}{2^{n-1}(1+\phi)}\left( \begin{array}{cc}
1 &  \text{WTr}(U_n^\dagger)   \\
\text{WTr}(U_n) &  1  
\end{array}
\right),
\]
which upon measurement of $\langle \sigma_x \rangle$ and $\langle \sigma_y \rangle$ yields the real and imaginary parts of $\mathcal{M} = \text{WTr}(U_n)/\bigl( 2^{n-1}(1+\phi)\bigr)$ respectively, where $n$ is the number of qubits in the bottom register.  The measured quantity $\mathcal{M}$ is then used to calculate the approximation of the Jones polynomial, $V(t)$, corresponding to the trace closure of the given braid at $t = e^{i2\pi/5}$,
\[
V(e^{i2\pi/5}) = (-(e^{i2\pi/5})^4)^{3w}\phi^{-1} \Big( 2^{n-1}(1+\phi) \mathcal{M} - \kappa \Big),
\]
where $\kappa = (2^{n-1} -f_m)\phi + (2^{n-1}- f_{m-1})$ and $w$ is the writhe of the braid, defined as the number of positive crossings minus the number of negative crossings.

Liquid state NMR offers one of the most advanced implementations of quantum information processors with high fidelity control of multiple qubits~\cite{Vandersypen2004NMR-Techniques-}.  The qubits are a bulk ensemble of identical spin-1/2 nuclei that exhibit a two-level energy structure in the presence of a strong magnetic field.  The ensemble of approximately $10^{20}$ molecules are manipulated in parallel and an ensemble measurement is performed using quadrature detection of the free induction decay to give $\langle\sigma_x\rangle$ and $\langle\sigma_y\rangle$.  The algorithm described above was demonstrated in liquid state NMR for the set of knots whose braid representations have four strands and three crossings.  There are six distinct knots in this set and hence, six distinct Jones polynomials.  The goal of the experiment is to distinguish between two distinct knots given their braid representations.  The subspaces of interest have $f_4 = 3$ and $f_{4-1} = 2$ basis states respectively, thus the encoding of the basis states requires $3$ qubits in the bottom register and a fourth as the control qubit.

\begin{figure}[htb]
\includegraphics[scale=0.08]{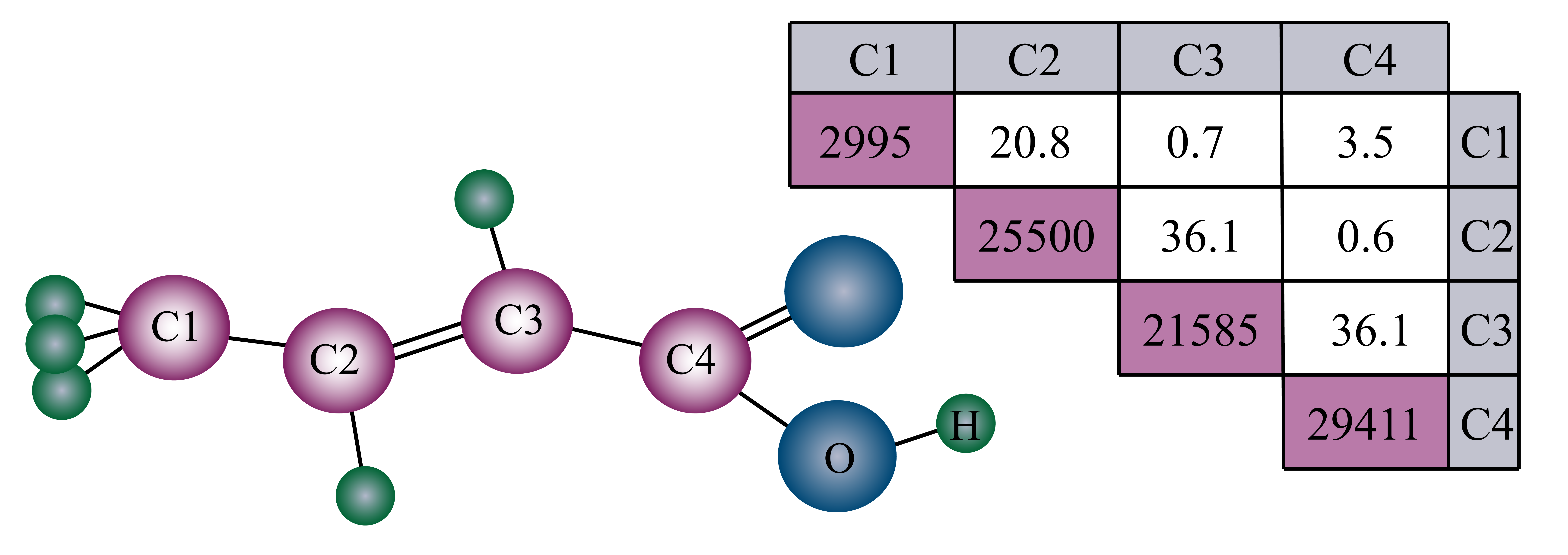}
\caption{The molecule trans-crotonic acid and a table with the parameters of the Hamiltonian given in Hz.  The shaded diagonal elements represent the chemical shifts $\omega_i$ with the Hamiltonian $\Sigma_i \pi\omega_i\sigma_z^i$.  The remaining elements indicate the scalar coupling constants $J_{ij}$ with the Hamiltonian $\Sigma_{i<j}\pi J_{ij}\sigma_z^i\sigma_z^j$.}
\label{molecule}
\end{figure}

The experiment was implemented on a Bruker Avance 700 MHz spectrometer using the molecule trans-crotonic acid (shown in fig \ref{molecule}).  The four qubits are experimentally realized by the four carbon nuclei, synthesized to be carbon-13, while the hydrogen are decoupled using the WALTZ-16~\cite{Shaka1983Evaluation-of-a} composite pulse sequence.   $C_1$ is our readout qubit whose initial state is the thermal state of $\rho = \openone + \varepsilon \mathbb{Z}$, $C_2$ is purified to the pseudo-pure state $\proj{0}$, and the remaining $C_3$ and $C_4$ are initialized to the completely mixed state.  
The radio frequency (r.f.) pulses that implement the unitary transformations are numerically generated using the GRAPE algorithm~\cite{Khaneja2005Optimal-control,RyanControl2008} which starts from a random guess and is then iteratively improved through a gradient ascent search.  
The GRAPE pulses are optimized to produce a fidelity $|\text{tr}(U^\dagger_{goal} U_{sim})|^2/d^2$, where $d$ is the dimension of the Hilbert space of $U_{goal}$, of no less than 0.998 and are designed to be robust to small inhomogeneities ($\pm 3\%$) in the r.f.~control field.  Each controlled-$\sigma_i$ unitary transformation is designed as a single pulse of 60 ms.  The pulses are corrected for non-linearities in the pulse generation and transmission to the sample by measuring the r.f. signal at the position of the sample using a feedback loop and iteratively modifying the pulse accordingly.  Through the feedback loop the implemented pulse can be measured and was found to have a simulated fidelity of 0.99 after correction.

\begin{figure}[h]
\includegraphics[scale=0.3]{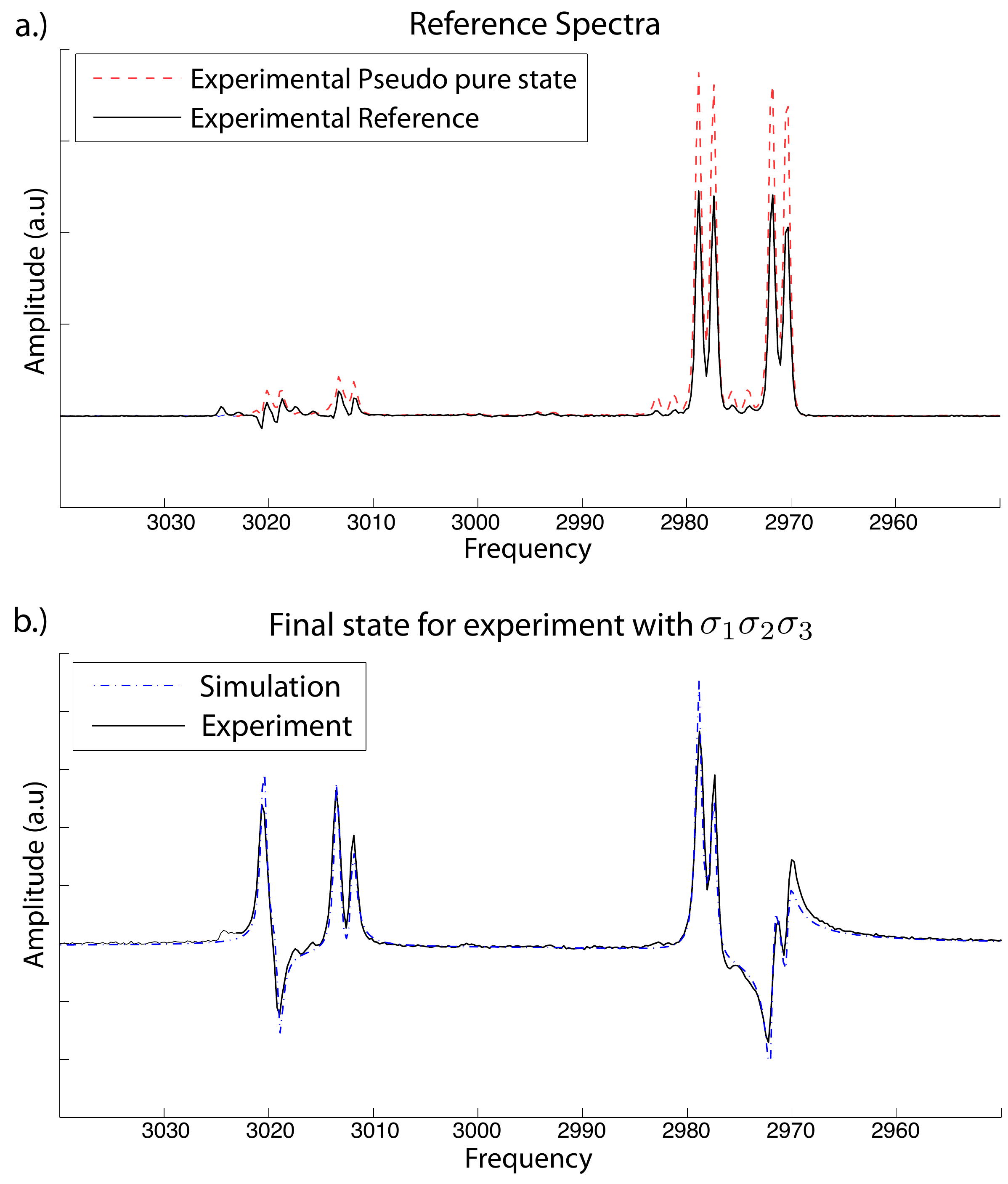}
\caption{(Color Online) The dashed (red) spectrum represents the pseudo pure state immediately after creation in the top (a) graph.  The solid (black) spectrum is the same pseudo pure state after 180 ms of pulses designed to perform the identity.  In the bottom graph (b), the solid spectrum (black) indicates the final state of the experiment and it is compared to simulation (dashed (blue)).  This particular experiment is for the knot whose braid representation has crossings $s_1s_2s_3$.}
\label{spectra}
\end{figure}

The resulting spectrum is fit and compared to a reference spectrum, traditionally of the initial state, to give the expectation value results.  In this experiment, pulses whose propagator was designed to be the identity were generated using GRAPE to have the same length and the same average power and fidelity as the controlled-$\sigma_i$.  These pulses were implemented and used to create a reference spectrum in an attempt to normalize some decoherence effects.  The state measured after three successive identity pulses, totaling 180 ms had only $60\%$ of the original signal (see figure \ref{spectra}), indicating this as a crucial step in the experimental procedure.  

\begin{figure}
\includegraphics[scale=0.51]{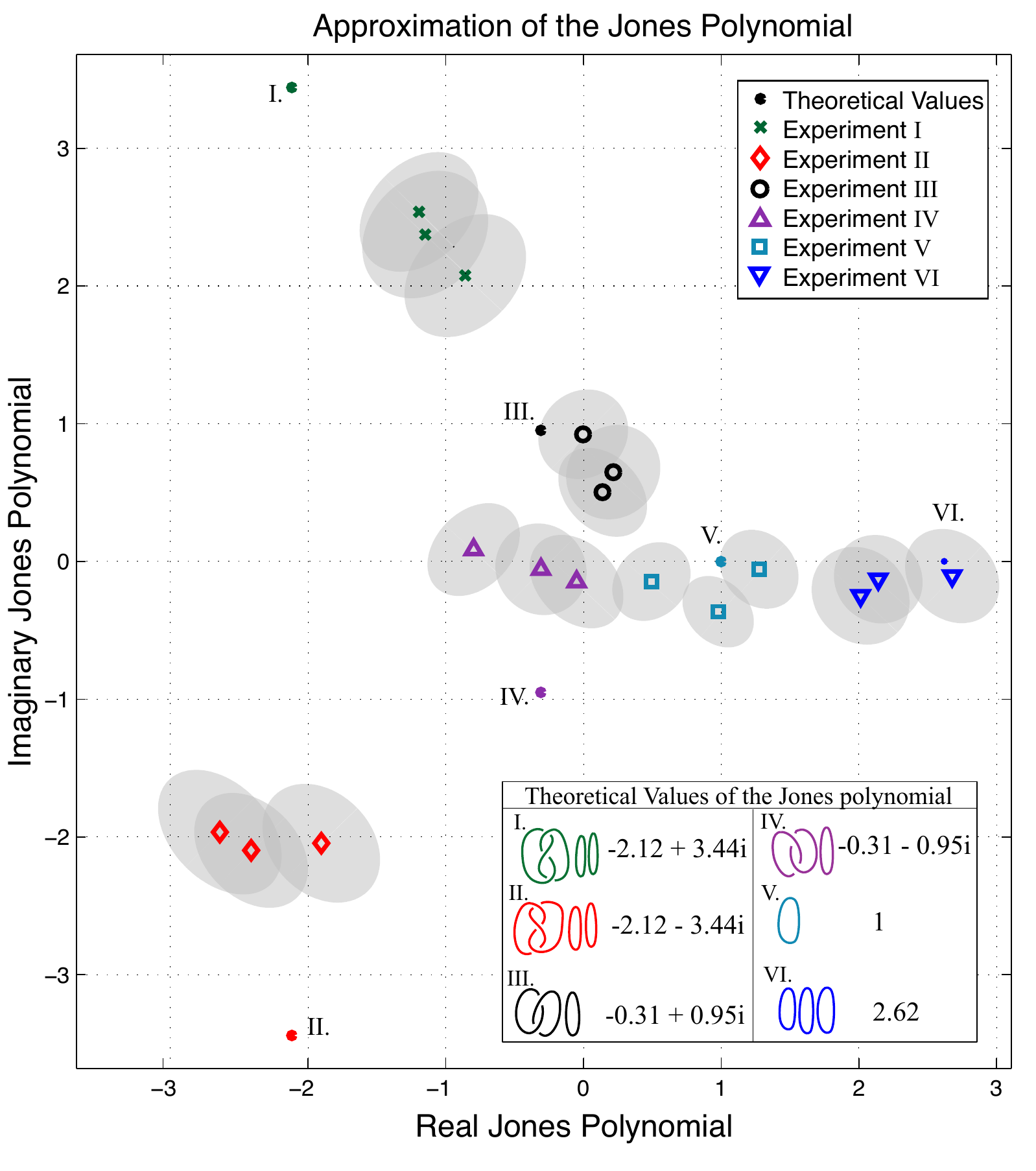}
\caption{The results for the approximation of the Jones polynomial for knots whose braid representations have 4 strands and 3 crossings.  There are six unique knots of this kind and their theoretical values of the Jones polynomial are plotted for the six experiments.  The corresponding experimental data points of three braid representations for each experiment are plotted along with error ellipses demonstrating the statistical error (with $86.5\%$ confidence levels or $2\sigma$).  The distribution is generated by simulating each experiment 200 times with single pulse fidelities of 0.99 which is the implemented pulse fidelity.  Using the error ellipses as discriminators, these results yields a $91\%$ success rate for distinguishing distinct knots.}
\label{results}
\end{figure}

The algorithm was implemented for 18 different braids, which correspond to 6 distinct Jones polynomials.  The results are displayed in figure \ref{results}.  
Systematic errors from imperfect initial state preparation and decoherence not captured by the reference state result in the offsets from the theoretical values.  The main contribution to the spreading of the experimental points is the finite fidelity of the optimal control pulses.

Two values of the Jones polynomial at best can distinguish between two knots if they are sufficiently far apart, and at worst, give no information, as even evaluations of the Jones polynomial that are identical would not be sufficient information to conclude the two knots are identical.  This leads to two types of errors when interpreting the data: \textit{passive} and \textit{fatal} errors.  Passive errors occur when two distinct knots are impossible to distinguish because of their relatively close distance to one another, while fatal errors occur when two identical knots are determined to be distinct.  The success rate for determining whether knots are distinct is calculated as the average of the percent of distinct knots correctly identified and the percent of identical knots correctly indistinguishable.
The error ellipses give a direct method for determining if two knots are distinct.  If the error ellipses for a pair of knots do not overlap then it is inferred that the knots are distinct, whereas if the two ellipses overlap no information is gained.  For the confidence region plotted in figure \ref{results}, 134 of the possible 135 pairs of distinct knots are correctly distinguished with 3 fatal errors of a possible 18, corresponding to a success rate of $91\%$.   

Approximation of the Jones polynomial is an example of a classical problem that appears intractable, but that can be solved using one clean qubit quantum computers.  This is the first experimental implementation of a DQC1-complete problem, and is performed in liquid state NMR with four qubits, resulting in a $91\%$ success rate for braids with four strands and a total of three crossings.  
In future work it will be interesting to see how the values of the Jones polynomial spread as you scale to larger knots and what size knot can be experimentally implemented before noise and control errors destroy the quantum advantage.

\begin{acknowledgments}
G.P. would like to thank M. Ditty for his technical expertise with the spectrometer.  This work was funded by NSERC, QuantumWorks, and CIFAR.  
\end{acknowledgments}


\end{document}